\documentclass[aps,prl,twocolumn,groupedaddress,showpacs]{revtex4-1}
\usepackage{amsmath}
\usepackage{relsize}
\usepackage{graphicx}
\usepackage{textcomp}
\newcommand {\ve} [1] {\mbox{\boldmath $#1$}}
\newcommand {\beq} {\begin{eqnarray}}
\newcommand {\eeqn} [1] {\label{#1} \end{eqnarray}}%
\newcommand {\eol} {\nonumber \\}

\begin{document}
\title{Sensitivity of (d,p) reactions to high n-p momenta and the consequences for nuclear spectroscopy studies}
\author{G.W. Bailey}
\affiliation{Department of Physics, Faculty of Engineering and Physical Sciences, University of Surrey
Guildford, Surrey GU2 7XH, United Kingdom}
\author{N.K. Timofeyuk}
\affiliation{Department of Physics, Faculty of Engineering and Physical Sciences, University of Surrey
Guildford, Surrey GU2 7XH, United Kingdom}
\author{J.A. Tostevin}
\affiliation{Department of Physics, Faculty of Engineering and Physical Sciences, University of Surrey
Guildford, Surrey GU2 7XH, United Kingdom}

\date{\today}
\begin{abstract}
Theoretical models of low-energy (d,p) single-neutron transfer reactions are a crucial link between experimentation, nuclear structure and nuclear astrophysical studies. Whereas reaction models that use local optical potentials are insensitive to short-range physics in the deuteron, we show that including the inherent nonlocality of the nucleon-target interactions and realistic deuteron wave functions generates significant sensitivity to high n-p relative momenta and to the underlying nucleon-nucleon interaction. We quantify this effect upon the deuteron channel distorting potentials within the framework of the adiabatic deuteron breakup model. The implications for calculated (d,p) cross sections and spectroscopic information deduced from experiments are discussed.
\end{abstract}
\maketitle

A universal feature of models of the nucleon-nucleon (NN) interaction is a strong repulsion at small NN separations. In atomic nuclei, this induces correlations between nucleon pairs at high relative momenta \cite{Dic04}, contributes to a reduction in the occupancies of nuclear single-particle states near the Fermi surfaces, and severely complicates practical computations of nuclear properties and observables. Modern theoretical nuclear physics methodologies attempt to transform this original, strongly-correlated many-body problem into one where the difficulties due to short-range repulsion are much reduced. A number of such approaches, that suppress explicit high-momentum components, have been proposed \cite{Bog02,Bog10} and applied \cite{Bog07} to model NN interactions -- such as derived from QCD-inspired chiral effective field theories (EFTs) \cite{Epe09,efts}.

Transforming away explicit high-momentum components through a {softening} of the NN interaction is often justified by assertions that low-energy observables are insensitive to these components \cite{Hol04,Jur08}. In this Letter we show that a very important class of nuclear reactions used for spectroscopic studies of nuclei, namely low-energy A(d,p)B reactions, can exhibit significant dependence on high n-p relative momenta. Specifically, this sensitivity is enhanced when including both the D-state component of the deuteron wave function, $\phi_0$, (the NN tensor force) and the inherent nonlocality of nucleon-nucleus optical potentials \cite{Fesh} in describing the deuteron-target (d-A) system in a model that accounts for deuteron breakup. This sensitivity challenges both the quantitative results and interpretation of spectroscopic studies of data from conventional, local A(d,p)B reaction model analyses.

\begin{figure}[t]
\includegraphics[scale=0.35]{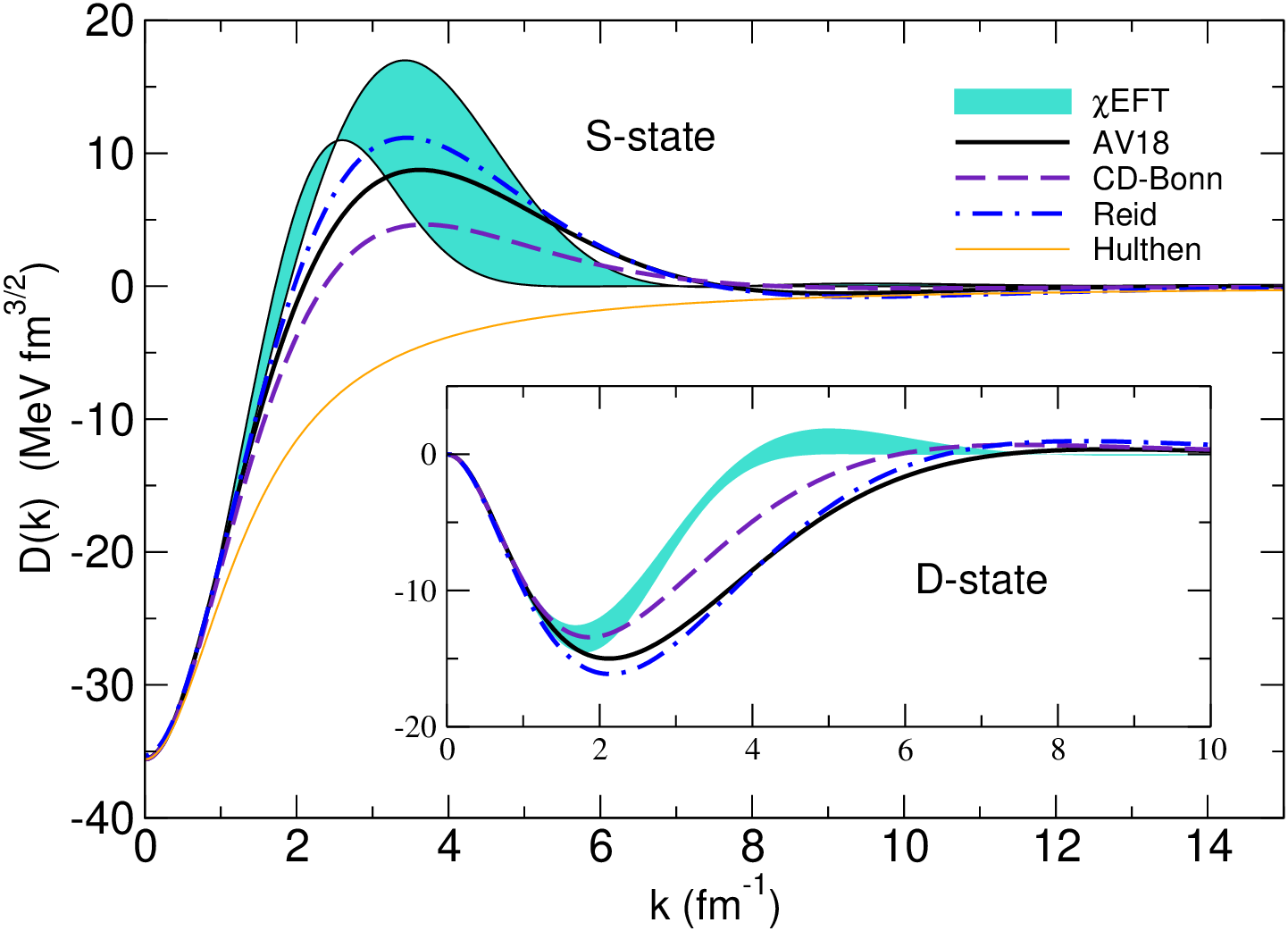}
\caption{Momentum space behaviors of the S- and D-state components of $D(\ve{k})= \langle\ve{k}|V_{np}| \phi_0\rangle$ for the NN potential models of Table I. The band for $\chi$EFT corresponds to the range of regulators in Table I.}
\label{fig:1}
\end{figure}

A strong indication of a possible high n-p momentum sensitivity of the A(d,p)B reaction was seen in a recent study \cite{TJprl,TJprc} that investigated the adiabatic model d-A potential, $U_{dA}$ \cite{Tan}, when including nonlocal nucleon-target (N-A) optical potentials. There, for $N=Z$ nuclei (and in leading order) it was shown that, if constructing $U_{dA}$ from local phenomenological n-A and p-A potentials, these potentials should be evaluated at an energy shifted by $\Delta E$ from that which is usually assumed, namely half the incident deuteron energy $E_d$. This energy shift was shown to be related to the following measure of the n-p relative kinetic energy, $T_{np}$, within the range of the n-p interaction $V_{np}$
%a property of the n-p system -- the expectation value of the n-p relative kinetic energy $T_{np}$ in the deuteron ground-state $\phi_0$ within the range of the n-p interaction $V_{np}$, weighted by the average n-p potential energy. Specifically, this value is
\begin{equation}
\langle T_{np}\rangle_V = \langle \phi_0|V_{np}T_{np}|\phi_0\rangle/\langle\phi_0|V_{np}|\phi_0\rangle\equiv\langle\phi_1|T_{np}|\phi_0\rangle ,
\nonumber
\end{equation}
where we have defined
%the deuteron vertex function,
$|\phi_1 \rangle = V_{np}|\phi_0 \rangle / \langle\phi_0 |V_{np} | \phi_0\rangle$. Determined by the properties of the ${^3}S_1- {^3}D_1$ NN channel at small n-p separations, and hence of $D(\ve{k}) = \langle\ve{k}|V_{np}| \phi_0\rangle$ at high n-p relative momenta, see Fig. \ref{fig:1}, the integrand of $\langle T_{np}\rangle_V$ is NN-model dependent. This NN model-dependence, driven principally by high n-p relative momenta, will affect A(d,p)B reaction observables.

The earlier $\Delta E $  value of Ref. \cite{TJprc} was obtained  assuming the purely attractive, phenomenological central Hulth\'en NN interaction and S-state wave function \cite{Hul}, whereas realistic deuteron wave functions have a modest D-state component with probability $P_D \approx 4-7\%$. Importantly however, for realistic model wave functions the matrix elements $\langle \phi_0|V_{np}| \phi_0\rangle$ entering $\langle T_{np}\rangle_V$ have D-state fractions ${\cal P}_D \approx 40\%$ (see Table I). The NN-model dependence and high n-p momentum components of $V_{np} \phi_0$ in Fig. \ref{fig:1} have implications for calculations of $U_{dA}$.

In this Letter we show that these high n-p momentum effects on $U_{dA}$ are considerably greater than is suggested by the modest $P_D$ values of realistic wave functions. We present exact calculations of the nonlocal adiabatic potential $U_{dA}$ from nonlocal nucleon-target optical potentials and realistic (S+D-state) deuteron wave functions. We quantify these effects of high n-p momenta on the local-equivalent potential to $U_{dA}$, and on calculated A(d,p)B cross-sections, and assess their potential impact on spectroscopic information deduced from transfer reaction data.

We discuss the A(d,p)B reaction in the context of the three-body reduction of the many-body transition amplitude \cite{FVA76,FVA78,PRP87}, which retains deuteron breakup effects, i.e.
\begin{equation}
T_{\rm (d,p)} = \sqrt{C^2S} \langle \chi_p^{(-)}\phi_n|V_{np}\,|\Psi_d^{(+)}\rangle .
\end{equation}
Here $\Psi_{d}^{(+)}(\ve{R},\ve{r})$ is the (three-body) wave function of the A+n+p system with an incident deuteron boundary condition, $\ve{r}$ and $\ve{R}$ are the n-p and d-A separations, and $\chi_p^{(-)}$ and $\phi_n$ are the distorted and bound wave functions of the proton and neutron in the final state. $C^2S$ is the spectroscopic factor. It has been shown \cite{Tan,Pang} that $T_{\rm (d,p)}$ converges very rapidly if $\Psi_{d}$ is expanded in the Weinberg states basis of the n-p system, and that $T_{\rm (d,p)}$ is well-described by retaining only the leading term. In this limit, the so-called adiabatic distorted waves approximation (ADWA), $\Psi_{d} (\ve{R}, \ve{r}) \rightarrow \chi_{dA} (\ve{R}) \phi_0(\ve{r})$, and the d-A distorting interaction $U_{dA}$ that generates the $\chi_{dA}$ is calculated from the n-A and p-A optical potentials using
\begin{eqnarray}
U_{dA} = \langle\phi_1|(U_{nA}+U_{pA})|\phi_0 \rangle.\label{JTeq}
\end{eqnarray}
Thus, $V_{np}$ enters $T_{\rm (d,p)}$ both: (i) explicitly, as the transition interaction in Eq. (1), and (ii) implicitly, within the adiabatic deuteron distorting potential $U_{dA}$, that generates the distorted waves $\chi_{dA}$.

We consider the following NN model descriptions: (i) the $S$-state Hulth\'en
%and Gaussian interactions \cite{Titus}
interaction \cite{Hul}, (ii) the phenomenological S+D-state Reid soft-core (RSC) \cite{RSC} and Argonne $v_{18}$ (AV18) \cite{v18} models, (iii) the meson-exchange CD-Bonn model \cite{CDB}, and (iv) very recent (N4LO) $\chi$EFT descriptions, for five different regulators \cite{efts}. In low-energy (and low momentum transfer) reactions, $T_{\rm (d,p)}$ is insensitive to the use of different model $V_{np}$ in the transition interaction. The S-state part of the transfer vertex $D(\ve{r}) = \langle\ve{r}|V_{np}| \phi_0\rangle$ enters via its volume integral, the zero-range normalization constant, $D_0$, and its finite-range parameter $\kappa$ \cite{Buttle}. The near-equality of these $D_0$ for the NN descriptions (i)--(iv) above is shown in Table I. Similarly, all range parameters $\kappa$ agree to within $4\%$. The D-state component of $D(\ve{r})$, quadratic in the n-p momentum for small momenta \cite{JoSan}, has a minimal effect on low energy (d,p) cross sections \cite{Del70,Knu75} and is not included. Our focus here is the sensitivity of $U_{dA}$ to the underlying NN model description, which then enters $T_{\rm (d,p)}$ through the $\chi_{dA}$

\begin{table}[h]
\centering
\begin{tabular}{ c | cc|cc|cc }
\hline\hline
NN Model &\ \ $P_D$\ \  & ${\cal P}_D$ & $D_0$  & $\kappa$  & $\langle\ T_{np}\rangle_V $ & $\Delta E$\\
& $\%$& $\%$&MeV fm$^{\frac{3}{2}}$&fm$^{-1}$&MeV&MeV\\
\hline
Hulth\'{e}n       &  0    & 0    &  $-$126.15  & 1.38 & 106.6 & 31.2 \\ %34.01 \\
%Gaussian          &  0    & 0    &  $-$126.87  & 2.48 & 49.1  & 30.78 \\
Reid soft core    & 6.46 &  39.7 &  $-$125.19  & 1.34 & 245.8 & 74.6 \\% 77.72 \\
Argonne V18       & 5.76 &  39.4 &  $-$126.11  & 1.32 & 218.0 & 66.2 \\%69.22 \\
CD-Bonn           & 4.85 &  32.6 &  $-$126.22  & 1.33 & 112.5 & 43.9 \\%46.74 \\
$\chi$EFT: 0.8 fm & 4.19 &  17.4 &  $-$126.17  & 1.34 & 247.2 & 71.6 \\%74.63 \\
$\chi$EFT: 0.9 fm & 4.29 &  19.7 &  $-$126.22  & 1.35 & 190.1 & 64.0 \\%66.98 \\
$\chi$EFT: 1.0 fm & 4.40 &  22.2 &  $-$126.32  & 1.36 & 154.6 & 57.0 \\%60.01 \\
$\chi$EFT: 1.1 fm & 4.74 &  26.1 &  $-$126.39  & 1.37 & 122.6 & 50.4 \\%53.38 \\
$\chi$EFT: 1.2 fm & 5.12 &  29.6 &  $-$126.50  & 1.38 & 88.2  & 44.2 \\%47.25 \\
\hline\hline
\end{tabular}
\caption{D-state percentages $P_D$ and ${\cal P}_D$ of $\langle\phi_0|\phi_0\rangle$ and
$\langle \phi_0 | V_{np} | \phi_0 \rangle$, respectively, volume integrals $D_0$ and
finite-range parameters $\kappa$ of the transfer vertex $D(\ve{r})$, and short-ranged
$n$-$p$ kinetic energy $\langle\ T_{np}\rangle_V $ for the different NN-model interactions
of the text. The $\chi$EFT wave functions use the different regulators shown. The energy
shifts $\Delta E$, calculated for the d+$^{26} $Al system at $E_d=12$ MeV, are computed
using the earlier lowest-order methodology of sections IV.A and IV.B of \cite{TJprc}.}
\label{tab:wfscomp}
\end{table}

In the following we will present results for $U_{dA}$ and (d,p) cross sections in the case of the $^{26} $Al(d,p)$^{27}$Al reaction, recently studied in connection with the destruction of $^{26}$Al in Wolf-Rayet and asymptotic giant branch stars \cite{Pai15,27Al}. Our calculations of $T_{\rm (d,p)}$ treat the transition interaction in zero-range approximation. The (d,p) reaction calculations are carried out using the transfer reactions code {\sc twofnr} \cite{twof}. We use the systematics of nonlocal nucleon potentials for $N=Z$ targets proposed by Giannini and Ricco (GR) \cite{GRicco} and a conventional parameter set, $r_0=1.25$ fm, $a_0=0.65$ fm, $V_{so}=6$ MeV, for the Woods-Saxon potentials that generate the neutron bound-state wave functions.

For orientation, we first consider the case when the nucleon optical potentials $U_{NA}$ are assumed local. $U_{dA}$ is then given by the Johnson-Tandy expression of Eq. (\ref{JTeq}). As the function $\phi_1$ is of short range, taking its zero-range limit yields the Johnson-Soper potential \cite{JSop}, $U_{dA}^{\rm JS}(R) =U_{nA}(R) +U_{pA}(R)$, the sum of the nucleon potentials evaluated at the deuteron centre-of-mass position. Thus, in this limit, $U_{dA}$ is independent of the NN interaction. This sensitivity returns when computing the full Johnson-Tandy expression but, for the realistic S+D-state deuteron wave function models, the depths of the calculated real and imaginary parts of $U_{dA}$ at the nuclear surface are found to differ by less than 1.5$\%$ and 3.7$\%$, respectively, for different realistic NN-model choices. The (d,p) cross sections, $\sigma_{\rm (d,p)}$, calculated using the same zero-range parameter $D_0$ to isolate the effects of changing the NN interaction within $U_{dA}$, changed by less than $0.6\%$. If we completely neglect deuteron breakup effects, by using the Watanabe model \cite{Wat}, then $\phi_1 \rightarrow \phi_0$ in Eq. (\ref{JTeq}) and there is further reduction in the sensitivity to the NN model choice as $U_{dA}^{\rm Wat}$ is now determined predominantly by the long-ranged parts of $\phi_0$ common to all models. The $\sigma_{\rm(d,p)}$ sensitivity to the NN models in this no-breakup limit is $\leq 0.4\%$.

Until very recently, all ADWA calculations have been performed assuming the $U_{NA}$ are local. We now calculate $T_{\rm (d,p)} $ assuming: (a) nonlocality of the nucleon-target potentials $U_{NA}$, and (b) the different NN model descriptions, including S+D-state deuteron wave functions, in constructing $U_{dA}$. The adiabatic deuteron distorting potential is now nonlocal, $U_{dA} (\ve{R},\ve{R}')$. This is calculated in the heavy target limit ($A \rightarrow \infty$) of Eq. (12) of Ref. \cite{TJprc}, the explicit expression for Eq. (\ref{JTeq}) when the nucleon potentials $U_{nA}$ and $U_{pA}$ are nonlocal. A new feature of these calculations is that the deuteron D-state is included. The deuteron distorted waves $\chi_{dA}$ now satisfy the integro-differential equation
\beq
(T_R + U_c(R) - E_d) \chi_{dA}(\ve{R}) &=& \eol - \int d\ve{R}' \,U_{dA}(\ve{R},\ve{R}') \chi_{dA}(\ve{R'}) &\equiv& -S(\ve{R}),
\eeqn{NLeq}
where $U_c(R)$ is the Coulomb interaction, assumed to act on the deuteron centre-of-mass. We use the analogous formula for the proton distorted waves $\chi_{p}$. The nonlocality of the $U_{NA}$ is taken to be of Perey-Buck form \cite{GRicco,PBuck}. The spin-orbit contribution is neglected. We solve Eq. (\ref{NLeq}) iteratively, after partial-wave expansion, and the numerical deuteron and proton channel partial-wave radial wave functions are read into the code {\sc twofnr} \cite{twof}. Full details of this formalism and procedure will be presented elsewhere \cite{long}. We note that, in the presence of the deuteron D-state, $U_{dA}$ includes off-diagonal (second-rank spin-tensor) terms. These are calculated but have negligible effect on the calculated (d,p) cross sections, so emphasis here is on the changes to the real and imaginary central terms of $U_{dA}$.

\begin{figure}[t]
\includegraphics[scale=0.28]{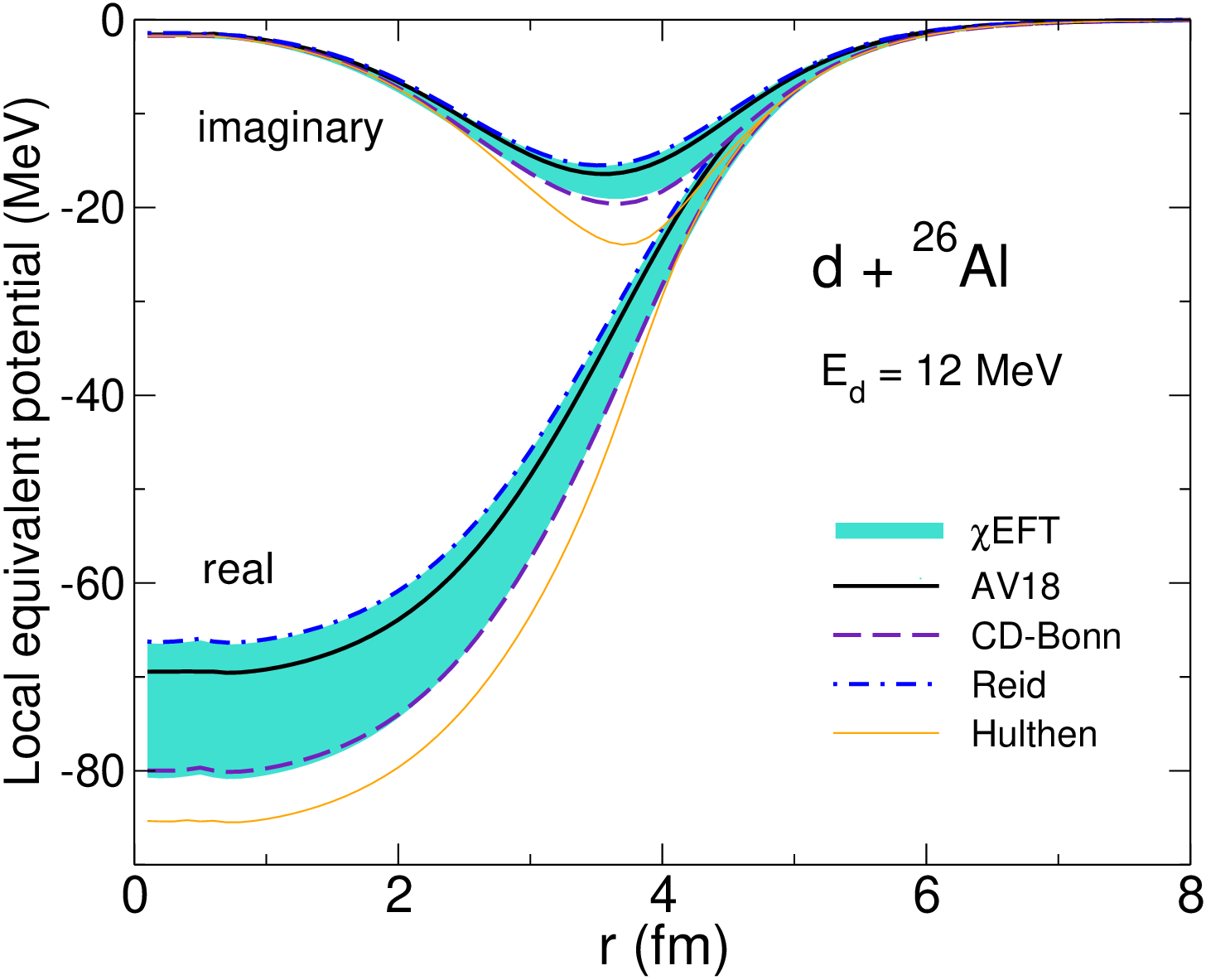}
\caption{Trivially-equivalent local deuteron-target potentials for the d+$^{26}$Al system at 12 MeV computed using the different NN-models of Table I. The band for $\chi$EFT corresponds to the range of regulators shown in Table I.}
\label{fig:2}
\end{figure}

The real and imaginary central terms of the trivially equivalent local potentials (TELP) to the nonlocal $U_{dA}(\ve{R},\ve{R}')$, defined as $S_l(R)/\chi_{dA,l}(R)$, are presented in Fig. \ref{fig:2} for the different NN-models of Table I. These TELPs are the same in each partial wave but their depths show a significant NN-model-dependence.
The essential features of these TELP to $U_{dA}$ from the all-order numerical calculations can be compared with the previously-reported local-equivalent description \cite{TJprc}. That approach, using a Taylor series expansion of the Perey-Buck nonlocal formfactor, can, in principle, be treated to arbitrary order. In leading order, the effects of the high n-p momenta enter as a shift $\Delta E$, to be added to $E_d/2$, for the energy at which the n-A and p-A potentials should be evaluated in constructing $U_{dA}$. The $\langle T_{np}\rangle_V$ values and the calculated $\Delta E$ values for the d+$^{26} $Al system at $E_d=12$ MeV (as in Ref. \cite{27Al}) from our different NN models are shown in Table I. $\langle T_{np}\rangle_V$ is maximal for the AV18 and RSC phenomenological potentials, being twice as large as for the CD-Bonn model. The values for the different $\chi$EFT models differ by almost a factor of three, and increase with decreasing value of the regulator radius. The values approach those of the phenomenological models for the smaller regulator radii. The corresponding energy shifts $\Delta E$ are $\sim 70$ MeV for the phenomenological NN models, compared to a shift of 31 MeV for the S-state only model. The result, when using realistic deuteron wave functions, is an increase in the energy shift $\Delta E$ by of order 10--40 MeV. For the energy dependence of the depth of the real GR  optical potential  this translates into a reduction in the real depth of the deuteron central potential of $\approx$ 2--18 MeV, consistent with the TELP changes in the full calculations, in Fig. \ref{fig:2}.

\begin{figure}[t]
\includegraphics[scale=0.32]{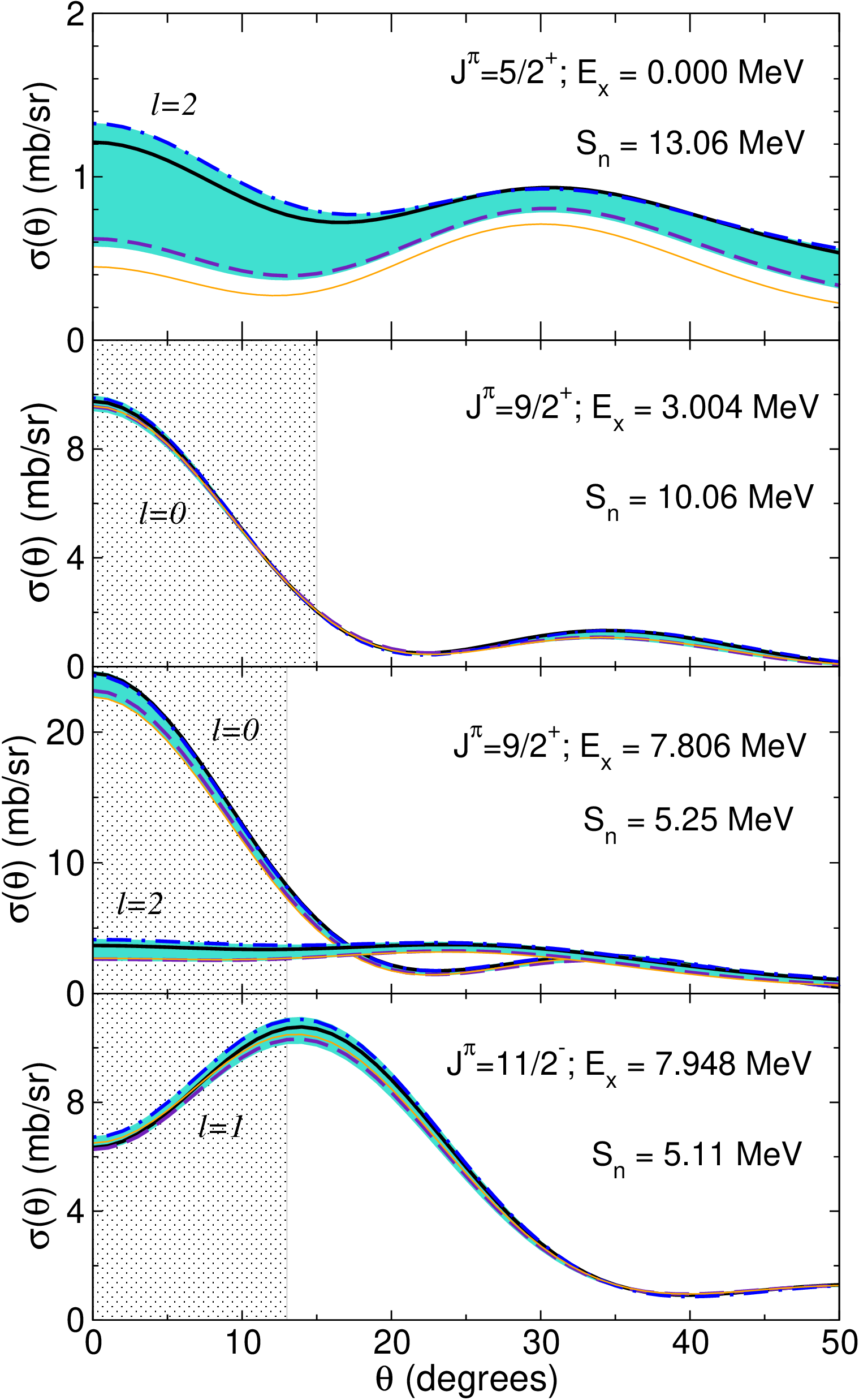}
\caption{Calculated differential cross sections for the $^{26}$Al(d,p) reaction at $E_d=12$ MeV for the four $^{27}$Al final states indicated and the different NN interaction models of Table I. The grey shaded areas cover the angular range of the experimental data reported in \cite{27Al}. All calculations use unit spectroscopic factors, $C^2S=1$. The different lines correspond to the same NN-models as in Figs. 1 and 2.}
\label{fig:3}
\end{figure}

The high n-p momentum sensitivity in the TELPs is reflected in the calculated
cross sections for $^{26}$Al(d,p)$^{27}$Al at $E_d = 12$ MeV, shown in Fig.
\ref{fig:3}, and depends on the transferred orbital momentum $\ell$ and neutron
separation energy to each $^{27}$Al final state. The effects for the $\ell=0,1$
transfers are found to be of order 5--10$\%$ in the angular region of the reported
experimental data. For the $\ell=2$ transfer component to the $J^\pi=9/2^+$ state
at $E_x = 7.806$ MeV - the analog of the astrophysically-important state in
$^{27}$Si - the effects on the cross section are 20--50$\%$, exceeding the
current experimental uncertainties, with implications for the deduced $\ell=0,2$
admixture and spectroscopic factors for this state. A fully quantitative analysis
of these effects on the deduced admixture and comparison with the results of
Ref. \cite{27Al}  will be presented in \cite{long}. An even stronger dependence, 
of 20--200$\%$, is observed for the $\ell=2$ transition to the $^{27}$Al ground 
state. Here the interpretation of experimental data would differ drastically from 
that using a conventional local potential analysis.

In summary, calculations of low-energy (d,p) reactions are presented that include both realistic deuteron wave functions and nonlocality of the nucleon-target optical potentials. We have demonstrated that calculations are sensitive to the NN-model used, through their different high n-p momentum content. This sensitivity only emerges when nonlocal optical nucleon potentials are used to construct the adiabatic deuteron channel potential, that accounts for deuteron breakup effects. The high n-p momenta in realistic S+D-state deuteron wave functions drive significant reductions to the depths of the deuteron-channel potential and, in some cases, increases in the calculated (d,p) cross sections. Here a Perey-Buck form was assumed for the nucleon nonlocality, but similar effects are anticipated if using nonlocal nucleon potentials derived from microscopic dynamical calculations. The results presented imply a significant uncertainty in conventional local (d,p) reactions analyses. Given the observed high momentum sensitivity, it would also be of interest to revisit the convergence properties of the Weinberg states expansion of the (d,p) transition amplitude in the presence of nucleon nonlocality.

We are grateful to E. Epelbaum for providing numerical values of the coordinate and momentum space N4LO $\chi$EFT and CD-Bonn deuteron wave functions. This work was supported by the United Kingdom Science and Technology Facilities Council (STFC) under Grant No. ST/L005743/1.

\bibliographystyle{apsrev4-1}

\begin{thebibliography}{50}
\bibitem{Dic04} W. Dickhoff and C. Barbieri, Prog. Part. Nucl. Phys. {\bf 52}, 377 (2004).
\bibitem{Bog02} S. Bogner, T.T.S. Kuo, L. Coraggio, A. Covello, and N. Itaco,
Phys. Rev. C {\bf 65}, 051301(R) (2002).
\bibitem{Bog10} S.K. Bogner, R.J. Furnstahl, and A. Schwenk, Prog. Part. Nucl. Phys.
{\bf 65}, 94 (2010).
\bibitem{Bog07} S.K. Bogner, R.J. Furnstahl, and R.J. Perry,
Phys. Rev. C {\bf 75}, 061001(R) (2007).
\bibitem{Epe09} E. Epelbaum, H.-W. Hammer, and Ulf-G. Mei{\ss}ner, Rev. Mod. Phys.
{\bf 81}, 1773 (2009).
\bibitem{efts} E. Epelbaum, H. Krebs and U.-G. Mei\ss{}ner, Phys. Rev. Lett. {\bf 115} 122301 (2015).
\bibitem{Hol04} J.D. Holt, T.T.S. Kuo, and G.E. Brown, Phys. Rev. C {\bf 69}, 034329 (2004).
\bibitem{Jur08} E.D. Jurgenson, S.K. Bogner, R.J. Furnstahl, and R.J. Perry,
Phys. Rev. C {\bf 78}, 014003 (2008).
\bibitem{Fesh} H. Feshbach, Ann. Rev. Nucl. {\bf 8}, 49 (1958).
\bibitem{TJprl} N.K. Timofeyuk and R.C. Johnson, Phys. Rev. Lett. {\bf 110}, 112501 (2013).
\bibitem{TJprc} N.K. Timofeyuk and R.C. Johnson, Phys. Rev. C. {\bf 87}, 064610 (2013).
\bibitem{Tan} R.C. Johnson and P.C. Tandy, Nucl. Phys. {\bf A235}, 56 (1974).
\bibitem{Hul} L. Hulth\'en and M. Sugawara, in Handbuch der Physik, Ed. by S. Fl�gge
(Springer-Verlag, Berlin, 1957), p. 1.
\bibitem{FVA76}
J.P Farrell Jr., C.M Vincent, N Austern, Ann. Phys. {\bf 96}, 333 (1976).
\bibitem{FVA78}
N Austern, C.M Vincent, J.P Farrell Jr., Ann. Phys. {\bf 114}, 93 (1978).
\bibitem{PRP87}
N. Austern, Y. Iseri, M. Kamimura, M. Kawai, G. Rawitscher, M. Yahiro,
Physics Reports {\bf 154}, 125 (1987).
\bibitem{Pang} D.Y. Pang, N.K. Timofeyuk, R.C. Johnson and J.A. Tostevin, Phys. Rev. C.
{\bf 87}, 064613 (2013).
%\bibitem{Titus} L.J. Titus, F.M. Nunes and G. Potel, Phys. Rev. C. {\bf 93}, 014604 (2016).
\bibitem{RSC} R.V. Reid, Ann. Phys. {\bf 50}, 411 (1968).
\bibitem{v18} R.B. Wiringa, V.G.J. Stoks and R. Schiavilla, Phys. Rev. C. {\bf 51}, 38 (1995).
\bibitem{CDB} R. Machleidt, Phys. Rev. C. {\bf 63} 024001 (2001).
\bibitem{Buttle} P.J.A Buttle and J.B Goldfarb, Proc. Phys. Soc. {\bf 83}, 701 (1964).
\bibitem{JoSan} R.C. Johnson and F.D. Santos, Phys. Rev. Lett. {\bf 19}, 364 (1967);
Part. Nucl. {\bf 2}, 285 (1971).
\bibitem{Del70}
G. Delic, B.A. Robson, Nucl. Phys. {\bf A156}, 97 (1970).
\bibitem{Knu75}
L.D. Knutson, J.A. Thomson, H.O. Meyer, Nucl. Phys. {\bf A241}, 36 (1975).
\bibitem{Pai15} S. D. Pain, D. W. Bardayan, J. C. Blackmon, S. M. Brown, K. Y. Chae, K. A. Chipps, J. A. Cizewski, K. L. Jones, R. L. Kozub, J. F. Liang, C. Matei, M. Matos, B. H. Moazen, C. D. Nesaraja, J. Okolowicz, P. D. O'Malley, W. A. Peters, S. T. Pittman, M. Ploszajczak, K. T. Schmitt, J. F. Shriner, Jr., D. Shapira, M. S. Smith, D. W. Stracener, and G. L. Wilson, Phys. Rev. Lett. {\bf 114}, 212501 (2015)
\bibitem{27Al} V. Margerin, G. Lotay, P. J. Woods, M. Aliotta, G. Christian, B. Davids,
T. Davinson, D.T. Doherty, J. Fallis,  D. Howell, O.S. Kirsebom, D.J. Mountford,
A. Rojas, C. Ruiz,  and J.A. Tostevin, Phys. Rev. Lett. {\bf 115}, 062701 (2015).
\bibitem{twof} J.A. Tostevin, University of Surrey version of the code {\sc twofnr}
(of M. Toyama, M. Igarashi and N. Kishida) and code {\sc front} (private communication).
\bibitem{GRicco} M.M. Giannini and G. Ricco, Ann. Phys. {\bf 102}, 458 (1976).
\bibitem{JSop} R.C. Johnson and P.J.R. Soper, Phys. Rev. C. {\bf 1} 3 (1970).
\bibitem{Wat} S. Watanabe, Nucl. Phys. {\bf 8}, 484 (1958).
\bibitem{PBuck} F. Perey and B. Buck, Nucl. Phys. {\bf 32}, 353 (1962).
\bibitem{long} G. W. Bailey, N. K. Timofeyuk and J.A. Tostevin, in preparation.
\end{thebibliography}

\end{document}